\documentclass[%
 reprint,
superscriptaddress,
 amsmath,amssymb,
 aps,
]{revtex4-2}

\usepackage{graphicx}% Include figure files
\usepackage{dcolumn}% Align table columns on decimal point
\usepackage{xcolor}
\usepackage{bm}% bold math
\begin{document}

\preprint{APS/123-QED}

\title{Contrasting Features of Parton Energy Loss  in Heavy-ion Collisions at RHIC and the LHC}

\author{Thomas Marshall}\email{rosstom@g.ucla.edu}
\affiliation{Department of Physics and Astronomy, University of
  California, Los Angeles, California 90095, USA}

\author{Philip Suh}
\affiliation{Department of Physics and Astronomy, University of
  California, Los Angeles, California 90095, USA}
\affiliation{Oxford Academy High School, Cypress, California 90630}
\affiliation{Stanford University, Stanford, California 94305, USA}
  
\author{Gang Wang}
\affiliation{Department of Physics and Astronomy, University of
  California, Los Angeles, California 90095, USA} 

\author{Huan Zhong Huang} \affiliation{Department of Physics and
  Astronomy, University of California, Los Angeles, California 90095,
  USA} \affiliation{Key Laboratory of Nuclear Physics and Ion-beam
  Application (MOE), and Institute of Modern Physics, Fudan
  University, Shanghai-200433, People’s Republic of China}

\date{\today}% It is always \today, today,

\begin{abstract}

Energetic quarks and gluons lose energy as they traverse the hot and dense medium created in high-energy heavy-ion collisions at the BNL Relativistic Heavy Ion Collider (RHIC) and the CERN Large Hadron Collider (LHC).
The nuclear modification factor ($R_{AA}$) of leading particles quantifies parton energy loss in such collisions, with the particle spectrum in $p+p$ collisions as a reference. Previous $R_{AA}$ measurements at RHIC energies have revealed an approximately constant trend at high transverse momenta ($p_{T}$), implying a scenario where parton energy loss, $\Delta p_{T}$, scales proportionally with $p_{T}$, a feature naively expected from energy loss dynamics in elastic collisions. In this study, we investigate the LHC $R_{AA}$ measurements which exhibit a pronounced $p_{T}$ dependence of $R_{AA}$ for various particle species, and our analysis attributes this behavior to $\Delta p_T$ being approximately proportional to $\sqrt{p_{T}}$. These distinct features are consistent with model calculations of dominant radiative energy loss dynamics at the LHC, in contrast to the dominance of collisional energy loss at RHIC. Additionally, the linear increase of the fractional energy loss with medium density at different $p_{T}$ magnitudes affirms our previous empirical observation that the magnitude of the energy loss depends more strongly on the initial entropy density rather than the parton's path length through the medium. Implications on the dynamical scenarios of parton energy loss and future experimental investigations will also be discussed.   

\begin{description}
\item[keywords]
heavy-ion collision; nuclear modification factor; parton energy loss
\end{description}
\end{abstract}

\maketitle

Color opacity stands as a fundamental trait of the hot and dense medium created in heavy-ion collisions at 
the BNL Relativistic Heavy Ion Collider (RHIC) and the CERN Large Hadron Collider (LHC). As energetic quarks and gluons traverse the medium, they shed energy through elastic scattering~\cite{PhysRevC.71.034907,PhysRevC.71.064904,PhysRevC.72.014905,PhysRevC.75.044906} and 
radiation of soft gluons~\cite{Dokshitzer:2001zm,ADIL200452,VITEV200638}.
In the scenario of an infinitely-high-momentum parton or an infinitely massive scattering center, energy loss would predominantly occur through radiative processes. Conversely, in the opposite scenario, collisional energy loss would become the dominant factor.
Prior empirical examinations of final-state leading particle spectra and the pertinent nuclear effects, using RHIC data, have revealed the proportionality between parton energy loss ($\Delta p_{T}$) and the magnitude of transverse momentum ($p_{T}$)~\cite{WangHuang:2008,PHENIX_WhitePaper,PHENIX_SLoss}. 
Although most theoretical energy loss calculations at RHIC are framed through the lens of radiative energy loss \cite{GLV1,Sadofyev_2021}, a proportional relationship between parton energy loss and transverse momentum tends to correspond better with a ``classical mechanics" understanding of elastic scattering and collisional energy loss, where higher-momentum bodies tend to lose proportionally more energy in collisions. Previous theoretical predictions additionally seem to show a linear dependence of the energy loss on the energy magnitude for heavy quarks, which would align with a similar dependence on $p_{T}$~\cite{mustafacollisional}.

Given that collision center-of-mass energies ($\sqrt{s_{NN}}$) at the LHC significantly surpass those at RHIC by over an order of magnitude, the associated $p_T$ range of generated particles now spans into a realm where radiative energy loss dynamics are expected to assume a more prominent role. 
Parton energy loss in the LHC center-of-mass energy regime has been previously studied through analysis of jet suppression $S_{\text{loss}}$ results with ATLAS \cite{ATLAS_SLoss}, but performing a similar study using leading particles instead allows for a cleaner look into the magnitude and the $p_{T}$ dependence of the energy loss without some compounding effects within jets. If energy loss and fragmentation were to occur sequentially—where fragmentation takes place outside the medium after the leading parton loses energy within the medium—leading particles would provide a clearer means of investigation than jets. Recent theoretical simulations of jet-induced medium excitation examine the diffusion wake in high-$p_T$ jet-photon pairs and predict an identical width, both with and without accounting for the medium, suggesting the possibility of separating energy loss and fragmentation~\cite{XNWJetMediumPaper}. 
Hence, the analysis of LHC data using the same framework as in Ref.~\cite{WangHuang:2008} is warranted to investigate the potential transition in the dynamics of energy loss from RHIC to the LHC through the study of leading particle data instead of jet data. 

Both radiative and collisional energy losses are intricately linked to the path length ($L$) and the entropy density of the medium.
We approximate the medium entropy density as $\frac{1}{S}\frac{dN}{dy}$, where $\frac{dN}{dy}$ represents the experimentally measured particle density per unit rapidity, and $S$ corresponds to the transverse overlap area of the colliding system, which can be determined using Monte Carlo Glauber calculations~\cite{PhysRevLett.86.3500,PhysRevC.65.031901,BEARDEN2001227,PhysRevC.79.034909}.
A previous study of RHIC data has unraveled a minimal dependence of $\Delta p_{T}$ on $L$, implying that 
parton energy loss is predominantly determined by 
the initial medium density~\cite{WangHuang:2008}. This feature could arise from the scenario of rapid expansion of the collision system, resulting in a swift decrease in medium entropy density over time. It is of great interest to investigate whether the LHC data corroborate the same characteristics.

In experiments, the nuclear modification factor, $R_{AA}$, quantifies the suppression or enhancement of particle yields in heavy-ion collisions relative to a nucleon-nucleon ($NN$) reference:
\begin{equation}
    R_{AA}(p_{T}) = \frac{d^{2}N^{AA}/dp_{T}d\eta}{T_{AA}d^{2}\sigma^{NN}/dp_{T}d\eta},
\end{equation}
where $T_{AA}$ accounts for the nuclear collision geometry, and $\eta$ denotes pseudorapidity. Both STAR \cite{STAR:2003fka}\cite{STAR:2006btx} and PHENIX \cite{PHENIX:2006wwy}\cite{Adare_2008} data demonstrate a plateauing of the $R_{AA}$ spectrum at values much lower than unity in the high-$p_{T}$ region ($\gtrsim$ 5 GeV/c). Treating the suppression of the nuclear modification factor as a result of empirical loss of transverse momentum from the $p$+$p$ spectrum to the nucleus+nucleus spectrum, these flat $R_{AA}$ curves were found to indicate a constant fractional $p_{T}$ shift in the spectrum. From a classical standpoint, this behavior is consistent with elastic collisional energy loss. Higher-$p_{T}$ particles would lose a proportionally higher amount of momentum through elastic collisions within the medium, resulting in a constant $\Delta p_{T}/p_{T}$. While this seems to describe the observed RHIC data fairly well, LHC data demonstrate significantly different characteristics.

\begin{figure}
\includegraphics[scale = 0.175]{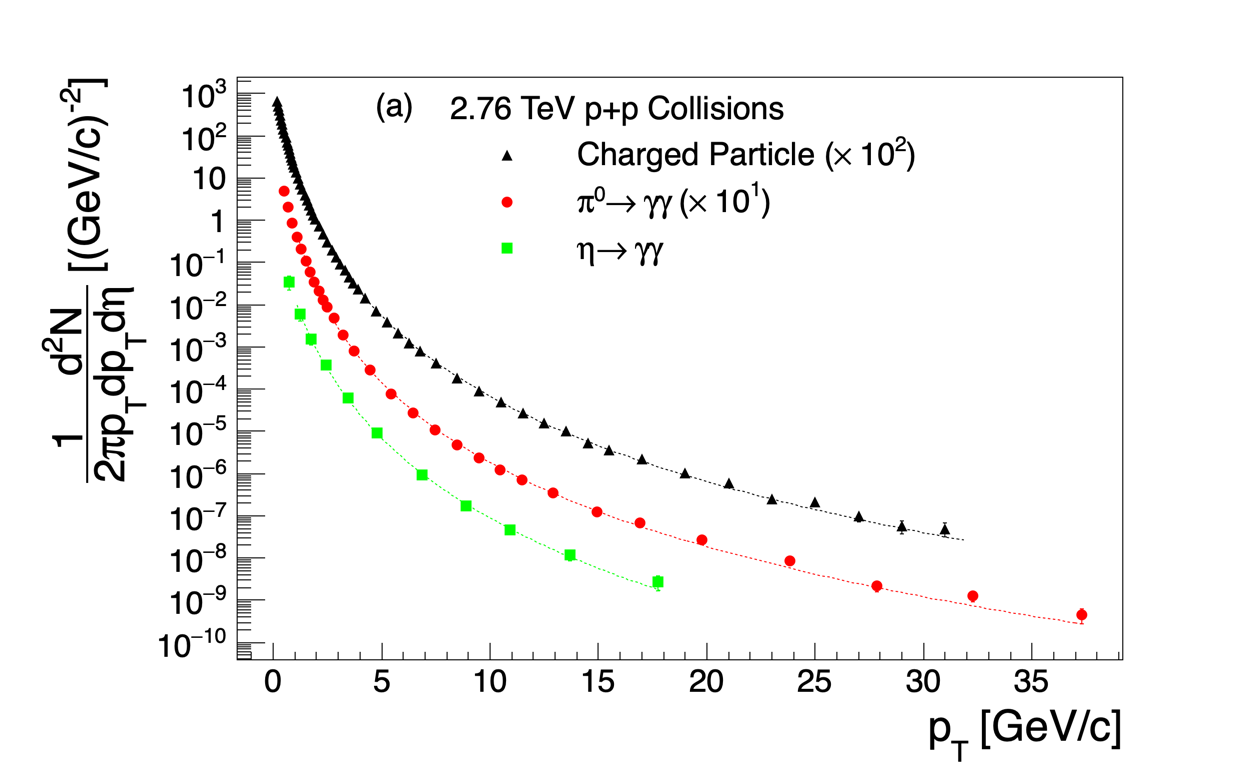}
\includegraphics[scale = 0.175]{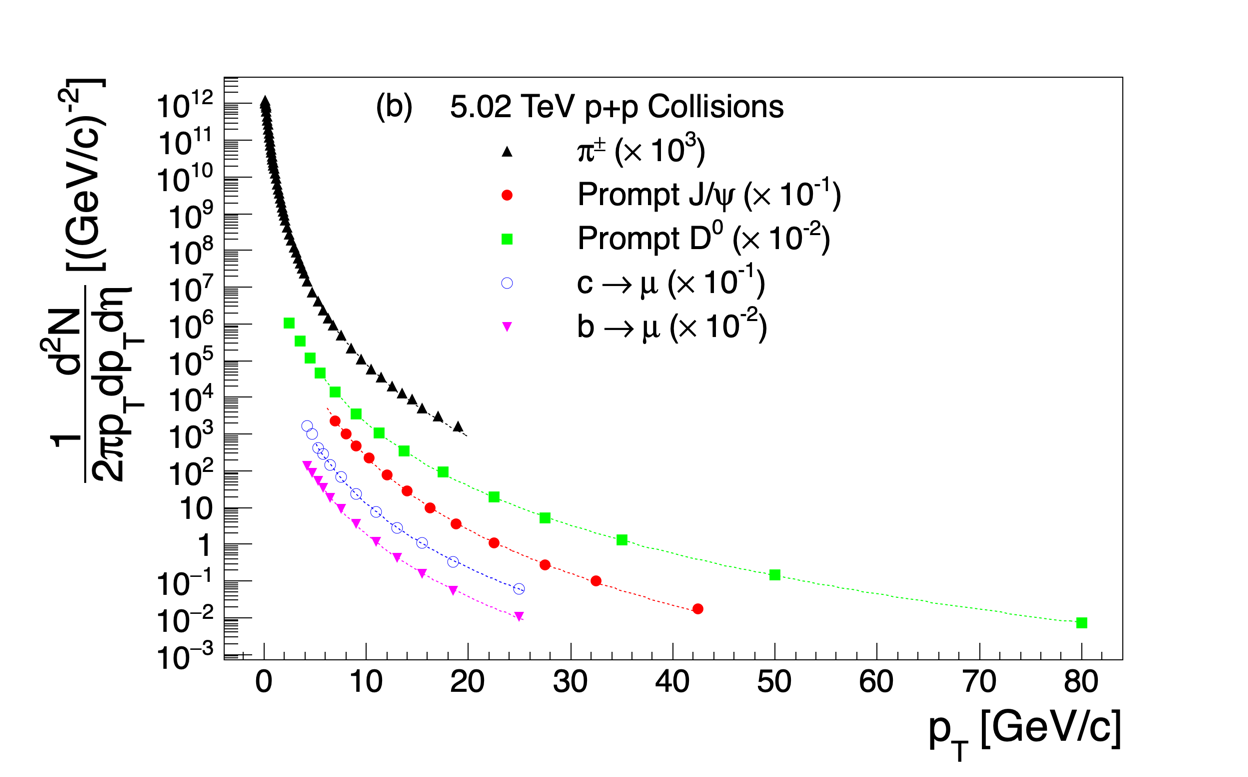}
\caption{Particle $p_{T}$ spectra in $p$+$p$ collisions at (a) 2.76 TeV and (b) 5.02 TeV. The 2.76 TeV data (charged particles, $\pi^{0}$, and $\eta$) are from ALICE~\cite{aliceChargedParticlePP,ALICEpionetaPP}.
The 5.02 TeV results include charged pions from ALICE~\cite{PhysRevC.101.044907}, prompt $J/\psi$ and prompt $D^{0}$ mesons from CMS~\cite{CMS:2017uuv,2018474}, and muons from charm and bottom hadrons from ATLAS~\cite{2022137077}. Different scaling factors are applied for better visibility. Note that some data sets used differential rapidity instead of pseudorapidity, but this should not affect the relevant physics involved in this study. Fits to the data follow Eq.~(\ref{eqn:powerlaw}) as discussed in the text.}
\label{fig:powerlaw}
\end{figure}

Figure~\ref{fig:powerlaw} depicts the published $p_T$ spectra of various final-state particles in $p$+$p$ collisions at (a) 2.76 TeV and (b) 5.02 TeV. Each dataset can be described by  a Tsallis distribution~\cite{TsallisFit}:
\begin{equation}
\frac{1}{2\pi p_{T}}\frac{d^{2}N}{dp_{T}d\eta}=A(1+\frac{p_{T}}{p_{0}})^{-n},
\label{eqn:powerlaw}
\end{equation}
where $A$, $p_{0}$, and $n$ are free parameters in the fit. The fit results for each of these parameters and the $\chi^{2}/$ndf (number of degrees of freedom) for each fit are listed in Tables \ref{tab:2.7TeVFitVals} and \ref{tab:5.02TeVFitVals}.

\begin{table}[bthp]
    \centering
\caption{Fit results for the $p_T$ spectra in $p$+$p$ collisions at $\sqrt{s_{NN}} =$ 2.76 TeV from Fig.~\ref{fig:powerlaw} using Eq.~(\ref{eqn:powerlaw}).}
    \begin{tabular}{c|c|c|c|c}
        Data Set & $A$ & $p_{0}$ & $n$ & $\chi^{2}/$ndf\\
        \hline
        charged particle & 4097 & 0.883 & 7.13 & 0.52\\
        $\pi^{0}\rightarrow\gamma\gamma$ & 312.0 & 0.696 & 6.94 & 0.82\\
        $\eta\rightarrow\gamma\gamma$ & 0.888 & 1.305 & 7.47 & 0.30
    \end{tabular}
    \label{tab:2.7TeVFitVals}
\end{table}

\begin{table}[bthp]
    \centering
    \caption{Fit results for the $p_T$ spectra in $p$+$p$ collisions at $\sqrt{s_{NN}} =$ 5.02 TeV from Fig.~\ref{fig:powerlaw} using Eq.~(\ref{eqn:powerlaw}).}
    \begin{tabular}{c|c|c|c|c}
        Data Set & $A$ & $p_{0}$ & $n$ & $\chi^{2}/$ndf\\
        \hline
        $\pi^{\pm}$ & $2.94\times10^{12}$ & 0.895 & 6.97 & 0.82\\
        prompt $J/\psi$ & $6.33\times10^{9}$ & 0.987 & 7.08 & 0.46\\
        prompt $D^{0}$ & $4.46\times10^{8}$ & 1.759 & 6.47 & 0.07 \\
        $c\rightarrow\mu$ & $5.41\times10^{7}$ & 0.957 & 6.25 & 0.13 \\
        $b\rightarrow\mu$ & $4.40\times10^{5}$ & 1.699 & 6.40 & 0.73
    \end{tabular}
    \label{tab:5.02TeVFitVals}
\end{table}

Following the procedures outlined in Ref.~\cite{WangHuang:2008} and treating the suppression empirically as a horizontal shift in the $p_{T}$ spectrum from $p$+$p$ to $A$+$A$ collisions, we can express $R_{AA}$ as
\begin{equation}
R_{AA}(p_{T})=\frac{(1+p'_{T}/p_{0})^{-n}p'_{T}}{(1+p_{T}/p_{0})^{-n}p_{T}}\left[1+\frac{dS(p_{T})}{dp_{T}}\right],
\label{eqn:RAAfit1}
\end{equation}
where $p'_{T} \equiv p_{T} + S(p_{T})$, and $S(p_{T})$ is the magnitude of the shift. 

The included factor of $1 + \frac{dS(p_T)}{dp_T}$ accounts for the necessary Jacobian term as mentioned in \cite{ATLAS_SLoss}.
Although $S(p_{T})$ being proportional to $p_T$ adequately describes RHIC data at the high-$p_T$ region, we start with a more general form in this paper, namely $S(p_{T}) = S_{0}{p_{T}}^{\alpha}$. Then, Eq.~(\ref{eqn:RAAfit1}) becomes
\begin{equation}
\begin{split}
    & R_{AA}(p_{T})=\frac{[1+(p_{T}+S_{0}{p_{T}}^{\alpha})/p_{0}]^{-n}(p_{T}+S_{0}{p_{T}}^{\alpha})}{(1+p_{T}/p_{0})^{-n}p_{T}}\\
    & \;\;\;\;\;\;\;\;\;\;\;\;\;\;\;\;\;\; \times (1+S_{0}\alpha{p_{T}}^{\alpha-1}).
    \label{eqn:RAAfit2}
\end{split}
\end{equation}
Once we determine $p_{0}$ and $n$ for each particle species from the $p_{T}$ distribution in Fig.~\ref{fig:powerlaw}, we regard them as fixed parameters in Eq.~(\ref{eqn:RAAfit2}), and use this formula to fit the corresponding $R_{AA}$ data allowing $S_{0}$ and $\alpha$ to vary as free parameters.

\begin{figure}[h!]
\centering
\includegraphics[scale=0.19]{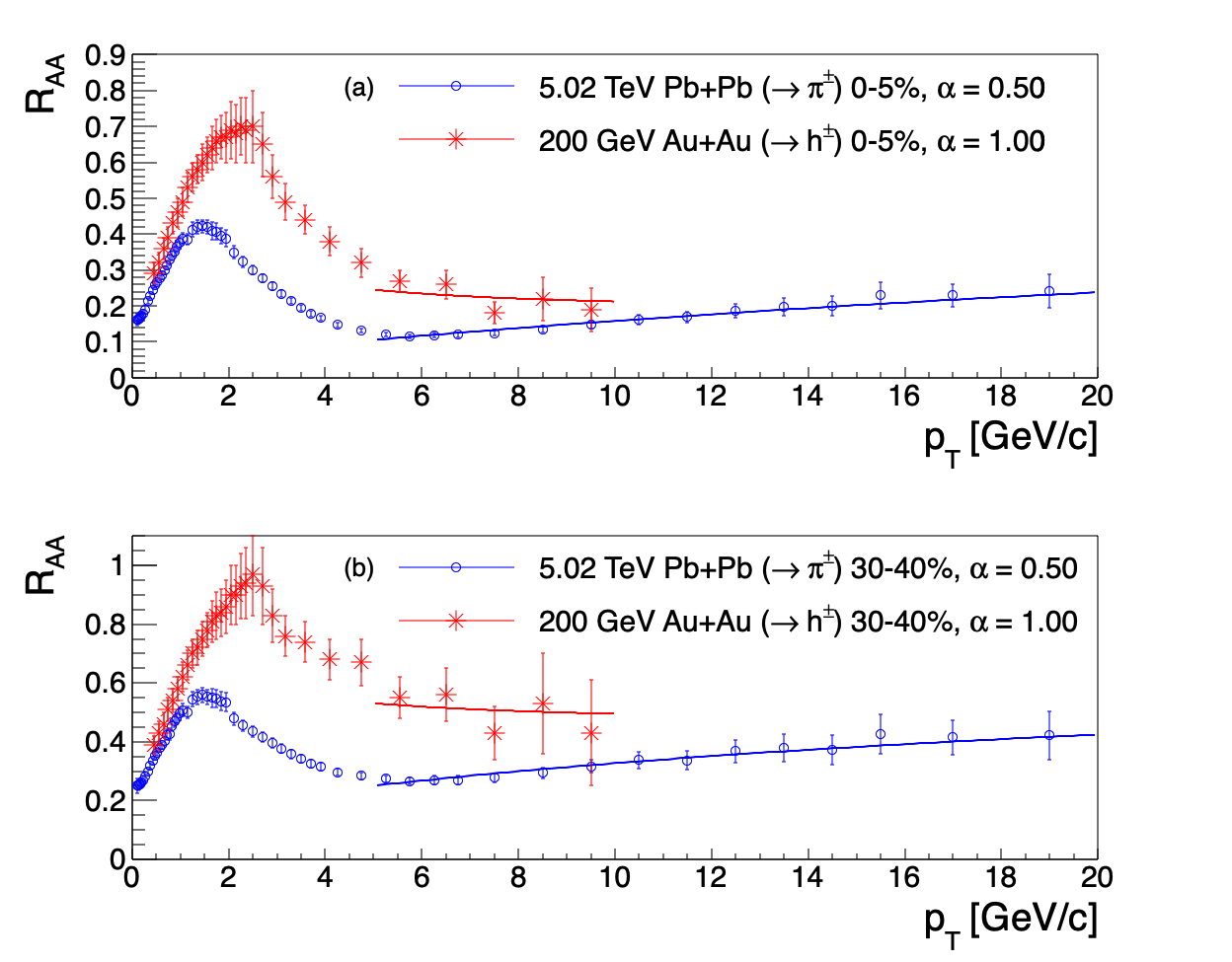}
\caption{$R_{AA}$ as a function of $p_T$ for charged hadrons in Au+Au collisions at 200 GeV (red)~\cite{STAR:2003fka} and for charged pions in Pb+Pb collisions at 5.02 TeV (blue)~\cite{PhysRevC.101.044907} for  (a) 0--5\% and (b) 30--40\% centrality ranges.
The fit functions follow Eq.~(\ref{eqn:RAAfit2}) with fixed $\alpha$ values of 1 and 0.5 for RHIC and the LHC data, respectively. The corresponding $p_{0}$ and $n$ values for the LHC data are extracted from the Tsallis fits in Fig.~\ref{fig:powerlaw}, and those for the RHIC data are taken from Ref.~\cite{WangHuang:2008}.}
\label{fig:raacomp}
\end{figure}

The necessity of introducing the $\alpha$ parameter is convincingly illustrated in Fig.~\ref{fig:raacomp}, which shows the $R_{AA}$ measurements as a function of $p_T$ for charged hadrons in Au+Au collisions at 200 GeV~\cite{STAR:2003fka} and for charged pions in Pb+Pb collisions at 5.02 TeV~\cite{PhysRevC.101.044907} for  (a) 0--5\% and (b) 30--40\% centrality ranges.
The fit functions adhere to Eq.~(\ref{eqn:RAAfit2}) with $S_0$ serving as the sole free parameter.
At $p_T \gtrsim$ 5 GeV/$c$, 
the flat $R_{AA}$ patterns at  RHIC agree with  $\alpha = 1$, whereas the increasing trends at the LHC harmonize with $\alpha = 0.5$. At both collision energies, the flattening and increasing trends initiate at approximately the same $p_{T}$ value of around 5 GeV/$c$. This pattern is also evident in the $R_{AA}$ data for other particle species to be presented, presumably because below this $p_T$ the soft physics dynamics including hydrodynamics and coalescence formation dominate, whereas above the $p_T$ of 5 GeV/$c$ parton fragmentation starts to dominate particle production where the parton energy loss picture emerges.

\begin{figure}
\centering
\includegraphics[scale=0.15]{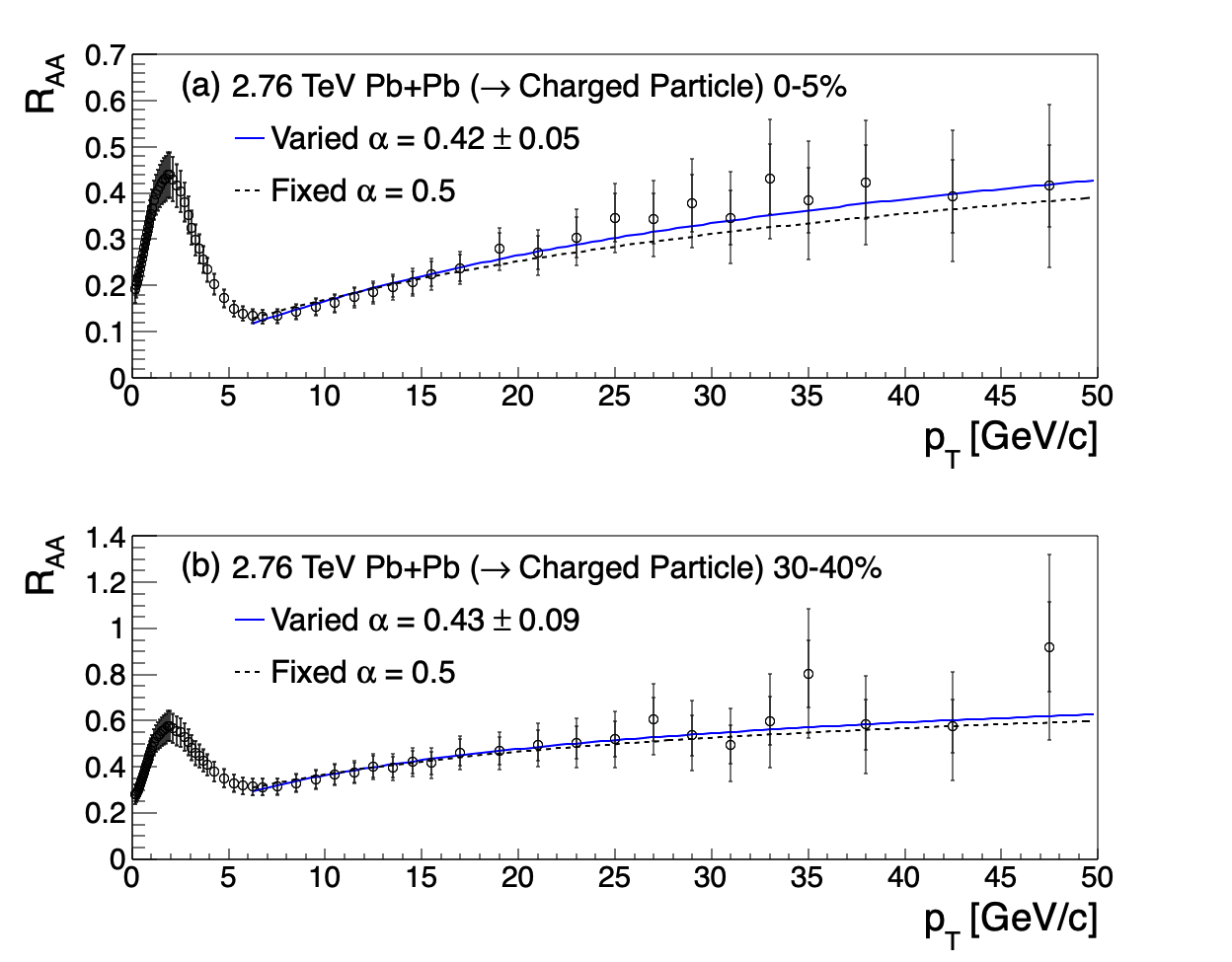}
\includegraphics[scale=0.15]{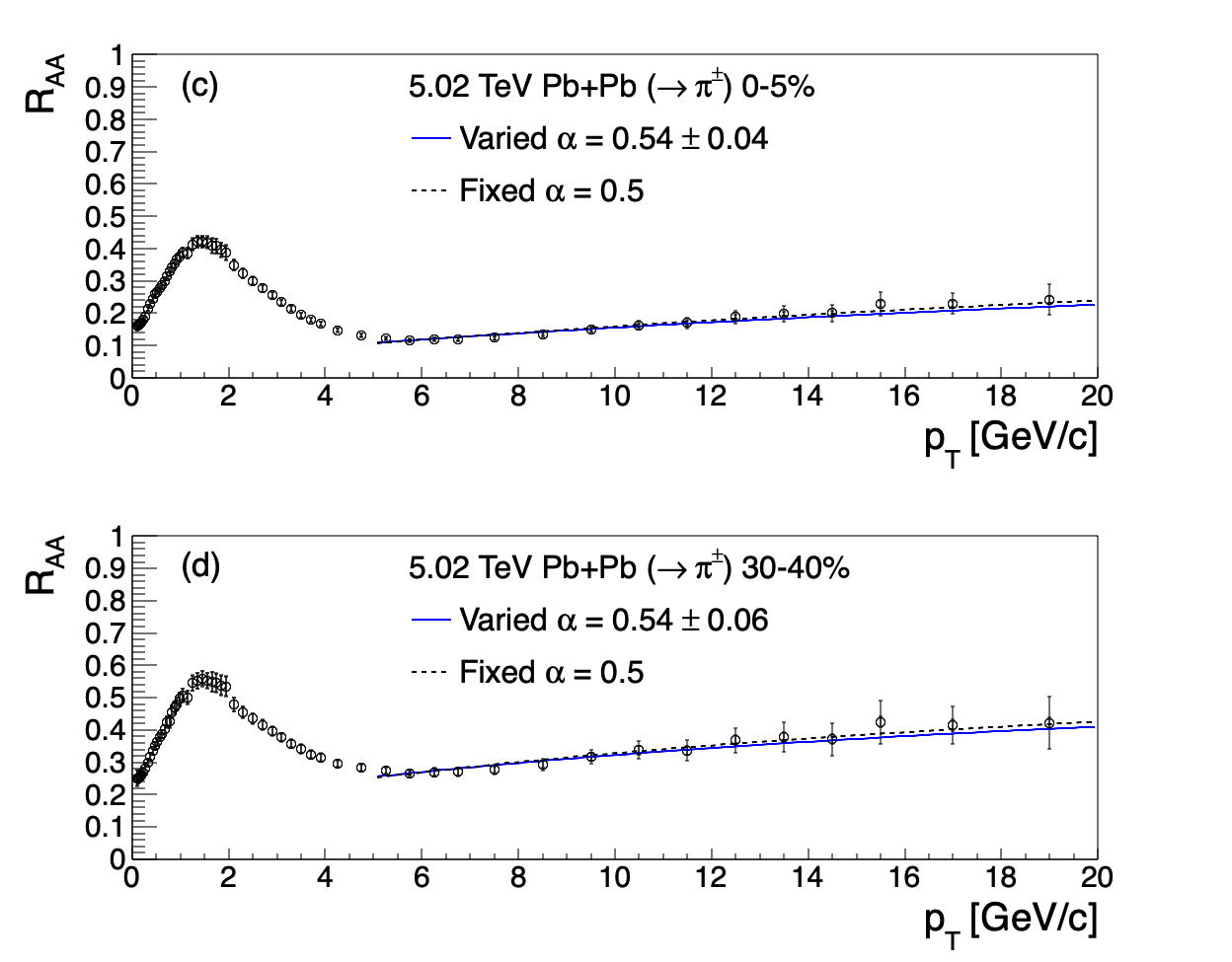}
\caption{$R_{AA}(p_T)$ for charged hadrons in (a) 0--5\%  and (b) 30--40\% Pb+Pb collisions at 2.76 TeV~\cite{ALICEChargedParticlePBPB}, and for charged pions in (c) 0--5\% and (d) 30--40\% Pb+Pb collisions at 5.02 TeV~\cite{PhysRevC.101.044907}. The fit functions from Eq.~(\ref{eqn:RAAfit2}) either take $\alpha$ as a free parameter or fix it at 0.5, using the $p_{0}$ and $n$ values extracted from the Tsallis fits in Fig.~\ref{fig:powerlaw}.}
\label{fig:raapion}
\end{figure}

\begin{table}[h!]
    \centering
\caption{Fit results for $R_{AA}(p_T)$ in Pb+Pb collisions at $\sqrt{s_{NN}} =$ 2.76 TeV from Figs.~\ref{fig:raapion} and \ref{fig:raacharm} using Eq.~(\ref{eqn:RAAfit2}).}
    \begin{tabular}{c|c|c|c}
        Data Set & $S_{0}$ & $\alpha$ & $\chi^{2}/$ndf\\
        \hline
        charged particle (0-5\%) & 1.56 & 0.42 & 0.18\\
        charged particle (30-40\%) & 0.80 & 0.43 & 0.51\\
        $\pi^{0}$ (0-10\%) & 1.44 & 0.50 & 0.11\\
        $\eta$ (0-10\%) & 1.73 & 0.41 & 0.08
    \end{tabular}
    \label{tab:2.76TeVRAAFitVals}
\end{table}

\begin{table}[h!]
    \centering
\caption{Fit results for $R_{AA}(p_T)$ in Pb+Pb collisions at $\sqrt{s_{NN}} =$ 5.02 TeV from Figs.~\ref{fig:raapion}, \ref{fig:raacharm}, and \ref{fig:raamuon} using Eq.~(\ref{eqn:RAAfit2}).}
    \begin{tabular}{c|c|c|c}
        Data Set & $S_{0}$ & $\alpha$ & $\chi^{2}/$ndf\\
        \hline
        $\pi^{\pm}$ (0-5\%) & 1.34 & 0.54 & 0.73\\
        $\pi^{\pm}$ (30-40\%) & 0.75 & 0.54 & 0.37\\
        prompt $J/\psi$ (0-100\%) & 0.76 & 0.54 & 0.30\\
        prompt $D^{0}$ (0-10\%) & 2.36 & 0.27 & 0.24\\
        $c\rightarrow\mu$ (0-10\%) & 0.48 & 0.83 & 0.20\\
        $b\rightarrow\mu$ (0-10\%) & 0.96 & 0.60 & 0.04
    \end{tabular}
    \label{tab:5.02TeVRAAFitVals}
\end{table}

Figure~\ref{fig:raapion} delineates $R_{AA}(p_T)$  for charged hadrons in (a) 0--5\%  and (b) 30--40\% Pb+Pb collisions at 2.76 TeV~\cite{ALICEChargedParticlePBPB}, and for charged pions in (c) 0--5\% and (d) 30--40\% Pb+Pb collisions at 5.02 TeV~\cite{PhysRevC.101.044907}.
All the datasets exhibit upward trends for $p_T \gtrsim$ 5 GeV/$c$. When we apply the same fitting approach and fix $\alpha$ at 0.5, the resulting fit curves (dashed lines) adequately capture all the data points.
When we take $\alpha$ as a free parameter (solid curve), 
the extracted $\alpha$ values are consistent with 0.5 within statistical uncertainties. The fit results for all the $R_{AA}$ studies performed in this study and the corresponding $\chi^{2}$/ndf are listed in Tables \ref{tab:2.76TeVRAAFitVals} and \ref{tab:5.02TeVRAAFitVals}.

\begin{figure}
\centering
\includegraphics[scale=0.15]{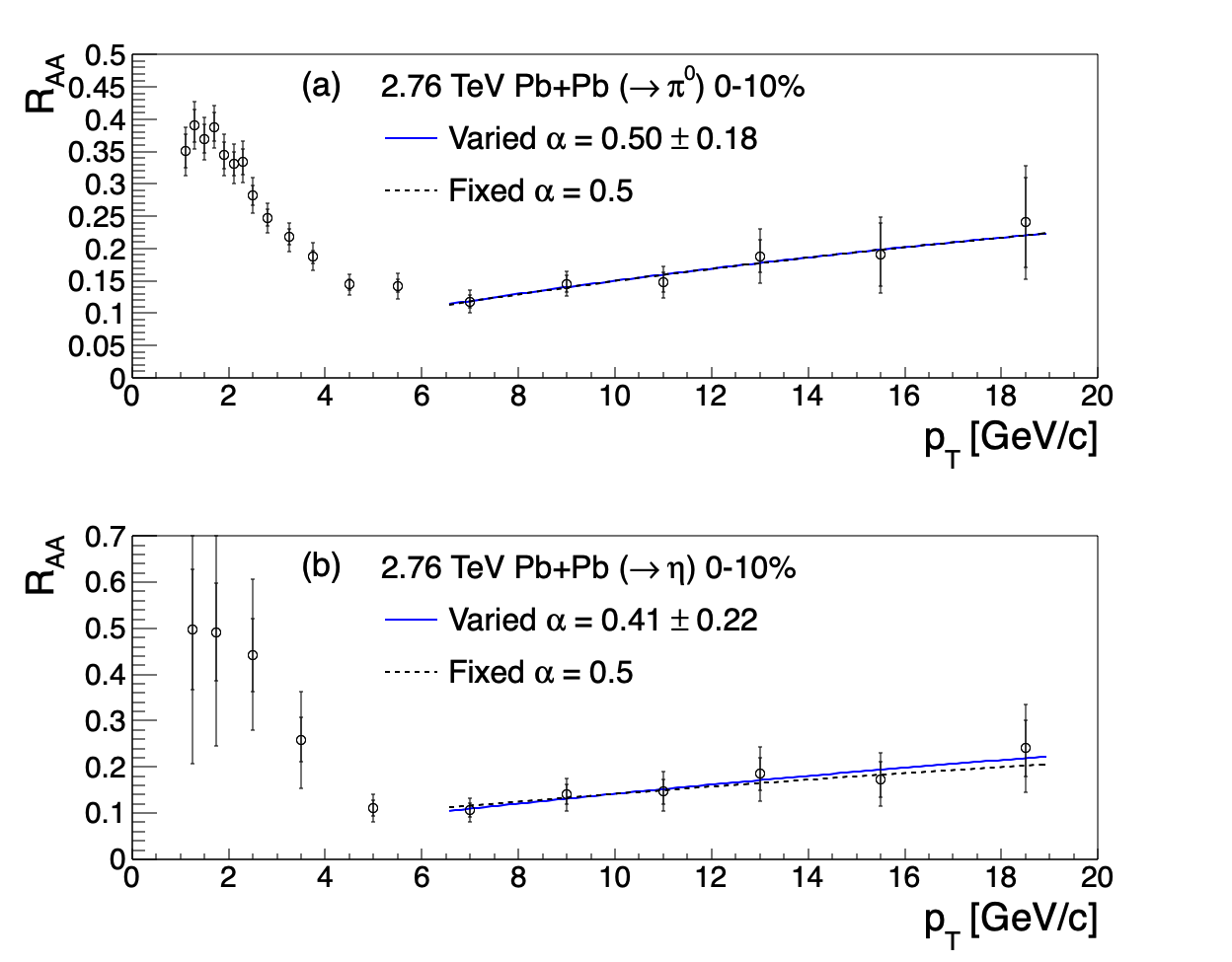}
\includegraphics[scale=0.15]{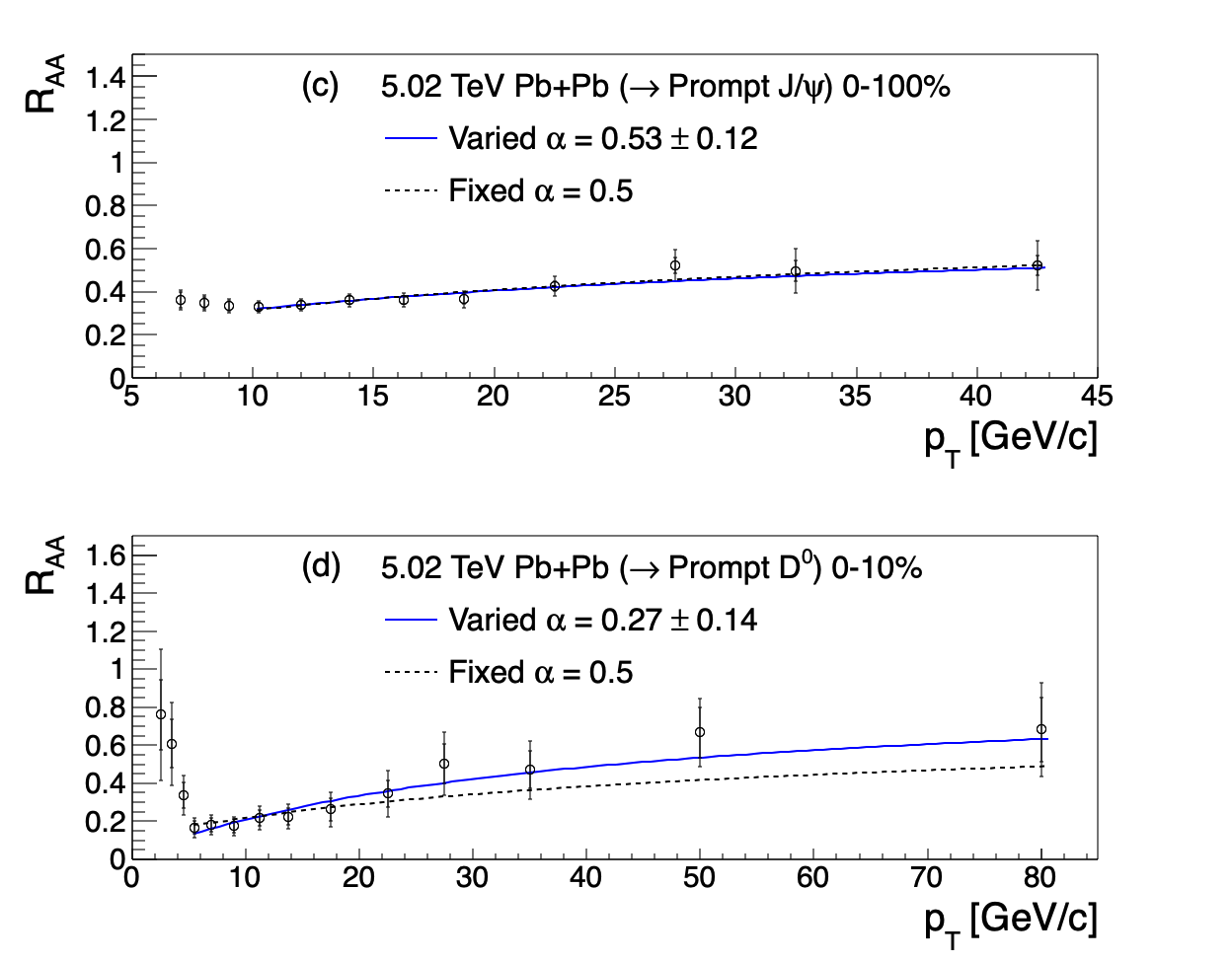}
\caption{$R_{AA}(p_T)$ for (a) $\pi^{0}$  and (b) $\eta$ mesons   in 0--10\%  Pb+Pb  collisions at 2.76 TeV~\cite{ALICEpionetaPbPb},  for (c) prompt $J/\psi$ mesons in  0--100\% Pb+Pb collisions at 5.02 TeV~\cite{CMS:2017uuv}, and for (d) prompt $D^{0}$ mesons in 0--10\% Pb+Pb collisions at 5.02 TeV~\cite{2018474}. The fit functions from Eq.~(\ref{eqn:RAAfit2}) either take $\alpha$ as a free parameter or fix it at 0.5, using the $p_{0}$ and $n$ values extracted from the Tsallis fits in Fig.~\ref{fig:powerlaw}.}
\label{fig:raacharm}
\end{figure}

We further investigate whether other final-state leading particles also exhibit these features. Figure~\ref{fig:raacharm} shows similar rising  trends of $R_{AA}$ at higher $p_{T}$ for (a) $\pi^0$ and (b) $\eta$ mesons in 0--10\% Pb+Pb at 2.76 TeV~\cite{ALICEpionetaPbPb}, for (c) prompt $J/\psi$ mesons in 0--100\% Pb+Pb at 5.02 TeV~\cite{CMS:2017uuv}, and for (d) prompt $D^{0}$ mesons in 0--10\% Pb+Pb at 5.02 TeV~\cite{2018474}. The $\alpha$ values extracted for  $\pi^{0}$, $\eta$, and $J/\psi$ mesons are consistent with 0.5 within the fitted statistical uncertainties. The fits to the prompt $D^{0}$ data seem to show some tension between the varied and fixed $\alpha$ values, but the significance of the deviation is only $1.6\sigma$. The fixed-parameter fit with $\alpha = 0.5$ agrees with nearly all the $D_{0}$ data points within uncertainties. Low precision beyond our $R_{AA}$ fit range in $p_{T}$ for RHIC data makes it difficult to determine whether the observed $\alpha = 0.5$ versus 1 behavior would also occur for heavy flavor leading particles. However, we would expect this to remain the case. Reference \cite{das2024charmhadronshothadronic} presents various transport model studies incorporating different energy loss mechanisms, which show good agreement with current RHIC and LHC data. Notably, the Duke model extends the standard Langevin equation by including medium-induced gluon radiation effects, successfully predicting $R_{AA}$ values consistent with observed $D$ meson suppression at the LHC. Future high-precision $D^0$ $R_{AA}$ measurements at high $p_{T}$ from both the LHC and RHIC will be crucial for further constraining the value of $\alpha$.

\begin{figure}
\centering
\includegraphics[scale=0.15]{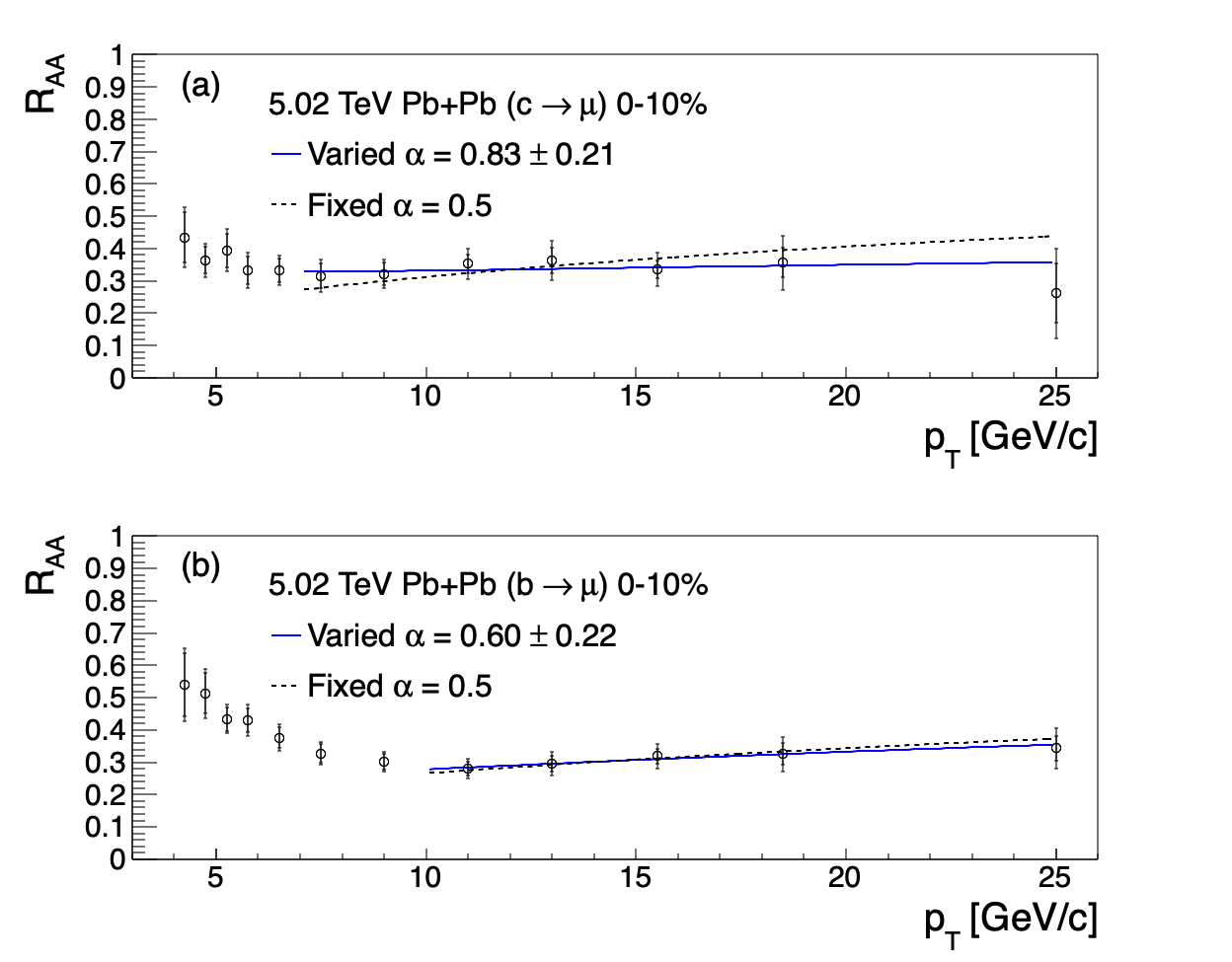}
\caption{$R_{AA}(p_T)$  for muons from (a) charm and (b) bottom hadrons in 0--10\% Pb+Pb collisions at 5.02 TeV~\cite{2022137077}. The fit functions from Eq.~(\ref{eqn:RAAfit2}) either take $\alpha$ as a free parameter or fix it at 0.5, using the $p_{0}$ and $n$ values extracted from the Tsallis fits in Fig.~\ref{fig:powerlaw}.}
\label{fig:raamuon}
\end{figure}

Figure~\ref{fig:raamuon} displays $R_{AA}(p_T)$ for muons originating from (a) charm and (b) bottom hadrons in 0--10\% Pb+Pb collisions at 5.02 TeV. In both cases, the fit curves with $\alpha$ = 0.5 align with all the data points within uncertainties. The $\alpha$ values extracted from the free-parameter fits exhibit a slight deviation from 0.5, with less than $1.5\sigma$ significance. Decay kinematics likely play a role in the smearing of the particle spectra for these cases, which may explain why the $\alpha$ values extracted for the muon data seem to trend above the 0.5 value we observe for the other data studied here.

To recap, the analyzed LHC data here suggest that to explain the $R_{AA}$ measurements for light- and heavy-quark hadrons as a $p_T$ shift in the spectrum from $p$+$p$ collisions, we require the corresponding  $\Delta p_T$ to scale with $\sqrt{p_{T}}$. This $p_T$ dependence contrasts with the previously observed proportionality with $p_{T}$ in RHIC data. Our analysis results with more recent data are in line with a previous study of LHC $R_{AA}$ data that determined the $\alpha$ value to be 0.55 \cite{Spousta_2016}. Our $p_T$ dependence of the parton energy loss at LHC supports theoretical predictions involving energy loss dynamics from medium-induced gluon radiation~\cite{baier2001quenching}. Thus, the distinct change from $\alpha=1$ at RHIC to $\alpha=0.5$ at LHC suggests a transition in the relative importance of collisional energy loss dynamics to radiative energy loss dynamics. 

This transition in the parton energy loss dynamics within the medium might also find an explanation in the effects of the jet-medium running coupling constant. Theoretical treatment of running coupling effects in both the radiative and elastic contributions to energy loss were previously shown to have robust agreement with LHC and RHIC data for light- and heavy-flavor $R_{AA}$ measurements \cite{Buzzatti_2013}. In this investigation, the emphasis was placed on the contributions from the radiated gluon vertex to the DGLV integral, particularly noting their significant role in enhancing the color transparency of the medium for higher-energy jets~\cite{Buzzatti_2013}. Should this apply to leading particles in addition to jets, our conclusion that radiative energy loss has an increased importance at the LHC relative to RHIC would be corroborated. While the limited ability to probe a broader $p_T$ spectrum at RHIC may affect the observation of the $p_T$ dependence of the strong coupling constant, we still observe a difference in the $R_{AA}$ trends within the same kinematic range. Within the $p_T \sim 5$--$10$ GeV/$c$ range in each collision system shown in Fig.~\ref{fig:raacomp}, a rising trend is evident in the LHC data but absent in the RHIC data. This observation suggests that the substantial difference in collision energy significantly contributes to this effect, independent of $p_T$-related effects on $\alpha$.

The dead-cone effect~\cite{DOKSHITZER2001199} predicts that gluon radiation is more strongly suppressed for bottom quarks than charm quarks, as the former bears a larger mass-to-energy ratio, leading to a wider dead cone. Recent measurements of heavy-quark meson production in $p$+$p$ collisions by the ALICE experiment~\cite{ALICE-deadcone-nature} reveal heavy-quark fragmentation in the vacuum and provide a direct observation of the dead-cone effect.
However, the LHC data of the $R_{AA}$ trends for muons from charm and bottom decays do not exhibit the anticipated reduced radiative energy loss for bottom quarks. The decay muon measurements can be influenced by various factors, including substantial momentum smearing resulting from decay kinematics, reduced sensitivity to low-momentum heavy-quark mesons, and the existence of non-prompt $c \rightarrow \mu$ decays that originate from $b$ quarks. Recent CMS~\cite{CMSNonPromptD0} and ALICE~\cite{ALICENonPromptD0} results indicate a significant reduction in the suppression of non-prompt charm hadrons over prompt charm hadrons, which could, in particular, contribute to the differences we see here via the latter factor. Another factor to consider is that the dead-cone effect may become less pronounced for very-high-energy quarks represented in the muon measurements. More precise data are needed to elucidate the nature of heavy-quark dynamics in the medium.  

\par In their similar study of $S_{\text{loss}}$, ATLAS additionally investigated the apparent differences in quark and gluon jet suppression through comparing inclusive and photon-tagged jet spectra~\cite{ATLAS_SLoss}. These differences were also explored in recent studies of the $p_{T}$ dispersion of inclusive jets and production cross sections of $J/\Psi$ mesons~\cite{Chen_2022,Song_2017}. While we do not explore this relationship in our study due to the challenging task of identifying whether our leading particle originates from a quark or gluon, it is crucial to underscore these prior findings and acknowledge their significance in the context of parton energy loss. 

\begin{figure}[b]
\centering
\includegraphics[scale=0.19]{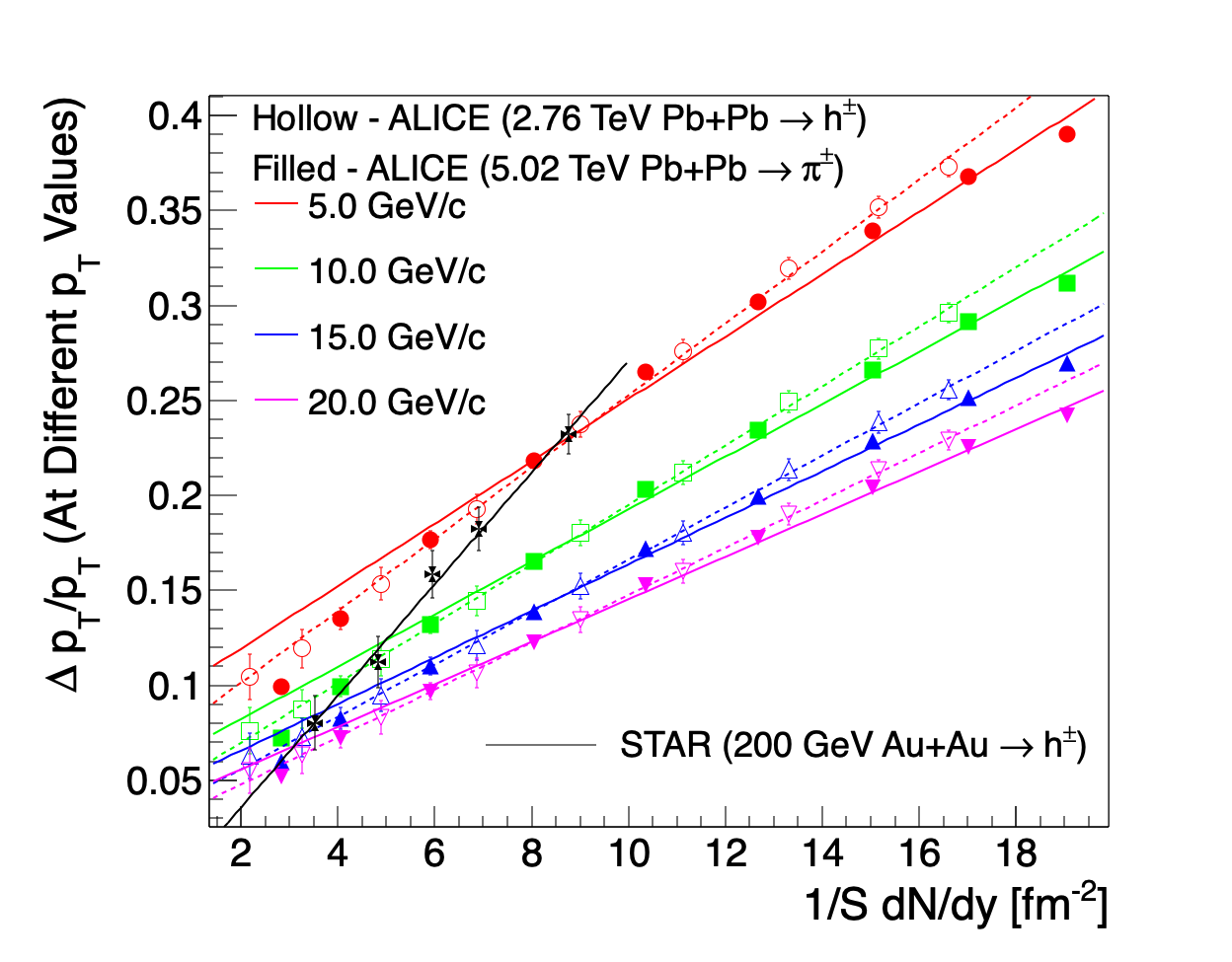}
\caption{Fractional energy loss ($\Delta p_{T}/p_{T}$) as a function of centrality (in terms of $\frac{1}{S}\frac{dN}{dy}$) based on the analysis in this study and Ref.~\cite{WangHuang:2008} for charged hadrons in Au+Au collisions at 200 GeV~\cite{STAR:2003fka} and in Pb+Pb at 2.76 TeV, and for charged pions in Pb+Pb at 5.02 TeV. Each of the LHC data points corresponds to a specific $p_{T}$ value used to calculate fractional energy loss, whereas fractional energy loss has no $p_{T}$-dependence in RHIC data~\cite{WangHuang:2008}. $\frac{1}{S}\frac{dN}{dy}$ values are calculated using the appropriate $S_{\perp}^{var}$ and $\frac{dN}{dy}$ columns from Table I of Ref.~\cite{PhysRevC.98.024904}. Each data set is fit separately with a linear function.}
\label{fig:pathLengthDep}
\end{figure}

We do, however, also investigate the relationship between energy loss and path length at LHC energies. Previous examinations of RHIC data have revealed that the deduced fractional energy loss, $\Delta p_{T}/p_{T}$, is a linear function of medium initial entropy density (quantified by $\frac{1}{S}\frac{dN}{dy}$) across different centrality intervals, despite significant variations in the path length for traversing partons~\cite{WangHuang:2008}. In contrast, when plotted as a function of path length (represented by various parameterizations using the number of participating nucleons $N_{\rm part}$), $\Delta p_{T}/p_{T}$ fails to exhibit a universal trend across different collision systems~\cite{WangHuang:2008}. This suggests that the initial collision density plays a more significant role in determining parton energy loss than the path length traversed by those partons. We apply the same analysis as adopted in Ref.~\cite{WangHuang:2008} to the LHC data and discover similar outcomes, as shown in Fig.~\ref{fig:pathLengthDep}. Now that fractional energy loss varies with $p_{T}$ according to the LHC $R_{AA}$ data, we plot $\Delta p_{T}/p_{T}$ as a function of $\frac{1}{S}\frac{dN}{dy}$ at different $p_T$ values for charged hadrons in Pb+Pb collisions at 2.76 TeV and for charged pions in Pb+Pb collisions at 5.02 TeV. In each case for each $p_{T}$ regime, a clear linear trend emerges, and the linearity is especially strong for higher $p_{T}$ scales, where parton fragmentation dominates particle production \cite{harris2023qgp}. 

\par We also find that the fractional energy loss decreases with increased transverse momentum at LHC energies, agreeing with the trend observed in previous ATLAS $S_{\text{loss}}$ measurements for inclusive jets \cite{ATLAS_SLoss}, as well as those found in Ref. \cite{harris2023qgp}. The linear trends between $\Delta p_{T}/p_{T}$ and $\frac{1}{S}\frac{dN}{dy}$ support the previous observation that the initial collision density has a stronger impact than path length, despite the significantly different initial densities at RHIC and the LHC. As discussed regarding RHIC data~\cite{WangHuang:2008}, a subdominant path-length dependence of energy loss might result from the rapid expansion of the medium, where most energy loss takes place before the parton traverses the full path length. Thereby, medium density becomes the dominant factor that determines the energy loss during the rapid expansion. Recent theoretical studies also suggest the formation time of partons could play a strong role in determining the medium induced energy loss~\cite{Zhang_2023}. We argue that in such a rapidly expansive medium, the static path length from the initial geometry of colliding nuclei fails to be the dominating factor for the parton energy loss in the medium. 

\par Due to limited statistics across multiple centralities, it is challenging to perform a similar measurement for the heavy-flavor hadrons studied here. Recent theoretical models incorporating both collisional and radiative energy loss mechanisms predict a constant value of $\Delta p_{T}/p_{T}$ beyond $p_{T} \sim 5$ GeV/$c$ for charmed hadrons in various collision systems at LHC energies~\cite{PhysRevC.111.014907}. This would disagree with the rising $p_{T}$ trends observed in the $D^{0}$ and $J/\psi$ $R_{AA}$ data in this study which point to $p_{T}$-dependent fractional energy loss in LHC data, unlike RHIC data. Although these trends are less prominent in the two muons from heavy quark $R_{AA}$ fits, we expect this is due to the decay kinematics of those measurements which may obscure the expected rising trend. Leading hadron measurements provide a much cleaner measurement, thus we argue that the predicted flat fractional energy loss trends from these models undervalue the weight of radiative energy loss which would likely produce the observed rising trends. These discrepancies highlight the importance of future heavy-flavor $R_{AA}$ measurements at sPHENIX and the LHC to investigate these effects further.

Our parton energy loss scenario predicts that the elliptic flow ($v_2$) of jets or high-$p_T$ leading particles in A+A collisions of intermediate centrality, if measured with minimal nonflow effects, will be close to zero and significantly smaller than values calculated assuming a static path-length dependence. Therefore, it is challenging to explain the large $v_2$ values observed for jets and leading particles at high $p_{T}$~\cite{ALICE_flow,Chen:2024aom}. Previous calculations of nuclear modification factor and anisotropic flow values for jets at RHIC and the LHC, however, do not prevent this possibility. In Ref.~\cite{PathLengthv2Calc}, a wide variety of energy loss models are shown to be able to describe available RHIC and LHC data within experimental uncertainty for both strong and weak path-length-dependent models. While this study indicates that the absence of path-length dependence is disfavored in the explored models inspired by quantum chromodynamics, it does not constrain the path-length-dependent parameter to be larger than zero~\cite{PathLengthv2Calc}. The exact degree of path-length dependence of parton energy loss remains an outstanding physics issue, which should be addressed with more precise data. It is puzzling why the observed $v_{2}$ for jets or leading particles shows a weak $p_T$ dependence, while the magnitude of parton energy loss exhibits a strong $p_T$ dependence. There are some experimental details worthy of more investigation such as the subtraction scheme for the $v_{2}$-dependent background. The nonflow correlations in elliptic flow measurements for jets and high-$p_T$ particles present an important factor to be accounted for as well. These experimental concerns should be adequately addressed before the path length dependence of the parton energy loss may be definitively determined.

In summary,  we present a parton energy loss study showing a significant distinction between RHIC and LHC data when empirically interpreting $R_{AA}(p_T)$ as a momentum loss in $A$+$A$ collisions relative to the $p$+$p$ reference. While the RHIC data favor a direct proportionality between the $p_{T}$ shift and $p_{T}$ itself, the LHC data suggest a proportionality with $\sqrt{p_{T}}$. This difference in the $p_T$ dependence signifies the heightened importance of radiative energy loss compared with collisional energy loss within the same transverse momentum range in colliding systems at higher $\sqrt{s_{NN}}$. Additionally, we find that the magnitude of the parton energy loss at LHC is largely determined by the initial medium entropy density, consistent with previous results at RHIC. This indicates that the path-length dependence of parton energy loss is less dominant, placing a greater emphasis on the initial medium density for a rapidly expanding medium. The distinct parton energy loss dynamics at RHIC and the LHC can be further investigated with high-statistics heavy-quark-tagged jets from the sPHENIX experiment at RHIC, as well as the LHC experiments in future runs. The possible sequential treatment of fragmentation and parton energy loss is of great interest for these future experiments. Equally interesting is the differentiation of energy loss contributions from quarks and gluons to the measurements studied here.

\begin{acknowledgments}
{The authors thank Dylan Neff, Jared Reiten, Dennis Perepelitsa, and Anthony Frawley for many fruitful discussions. T. M., P. S., G. W., and H. H. are supported
by the U.S. Department of Energy under Grant No. DE-FG02-88ER40424 and by the National Natural Science Foundation of China under Contract No.1835002.
}
\end{acknowledgments}

\bibliographystyle{JHEP}
\bibliography{ref.bib}

\begin{thebibliography}{10}

\bibitem{PhysRevC.71.034907}
H.~van Hees and R.~Rapp.
\newblock Phys. Rev. C, \textbf{71}:034907 (2005)

\bibitem{PhysRevC.71.064904}
G.~D. Moore and D.~Teaney.
\newblock Phys. Rev. C, \textbf{71}:064904 (2005)

\bibitem{PhysRevC.72.014905}
M.~G. Mustafa.
\newblock Phys. Rev. C, \textbf{72}:014905 (2005)

\bibitem{PhysRevC.75.044906}
A.~Adil, M.~Gyulassy, W.~Horowitz, \emph{et~al.}
\newblock Phys. Rev. C, \textbf{75}:044906 (2007)

\bibitem{Dokshitzer:2001zm}
Y.~L. Dokshitzer and D.~E. Kharzeev.
\newblock Phys. Lett. B, \textbf{519}:199--206 (2001)

\bibitem{ADIL200452}
A.~Adil and M.~Gyulassy.
\newblock Physics Letters B, \textbf{602}~(1):52--59 (2004).
\newblock ISSN 0370-2693

\bibitem{VITEV200638}
I.~Vitev.
\newblock Physics Letters B, \textbf{639}~(1):38--45 (2006).
\newblock ISSN 0370-2693

\bibitem{WangHuang:2008}
G.~Wang and H.~Z. Huang.
\newblock Physics Letters B, \textbf{672}~(1):30--34 (2009).
\newblock ISSN 0370-2693

\bibitem{PHENIX_WhitePaper}
K.~Adcox, S.~Adler, S.~Afanasiev, \emph{et~al.}
\newblock Nuclear Physics A, \textbf{757}~(1–2):184–283 (2005).
\newblock ISSN 0375-9474

\bibitem{PHENIX_SLoss}
A.~Adare, S.~Afanasiev, C.~Aidala, \emph{et~al.}
\newblock Physical Review C, \textbf{93}~(2) (2016).
\newblock ISSN 2469-9993

\bibitem{GLV1}
M.~Gyulassy, P.~Levai, and I.~Vitev.
\newblock Nuclear Physics B, \textbf{594}~(1):371--419 (2001).
\newblock ISSN 0550-3213

\bibitem{Sadofyev_2021}
A.~V. Sadofyev, M.~D. Sievert, and I.~Vitev.
\newblock Physical Review D, \textbf{104}~(9):094044 (2021).
\newblock ISSN 2470-0029

\bibitem{mustafacollisional}
M.~G. Mustafa.
\newblock Physical Review C—Nuclear Physics, \textbf{72}~(1):014905 (2005)

\bibitem{ATLAS_SLoss}
G.~Aad, B.~Abbott, K.~Abeling, \emph{et~al.}
\newblock Physics Letters B, \textbf{846}:138154 (2023).
\newblock ISSN 0370-2693

\bibitem{XNWJetMediumPaper}
W.~Chen, S.~Cao, T.~Luo, \emph{et~al.}
\newblock Phys. Lett. B, \textbf{777}:86--90 (2018)

\bibitem{PhysRevLett.86.3500}
K.~Adcox, S.~S. Adler, N.~N. Ajitanand, \emph{et~al.}
\newblock Phys. Rev. Lett., \textbf{86}:3500--3505 (2001)

\bibitem{PhysRevC.65.031901}
B.~B. Back, M.~D. Baker, D.~S. Barton, \emph{et~al.}
\newblock Phys. Rev. C, \textbf{65}:031901 (2002)

\bibitem{BEARDEN2001227}
I.~Bearden, D.~Beavis, C.~Besliu, \emph{et~al.}
\newblock Physics Letters B, \textbf{523}~(3):227--233 (2001).
\newblock ISSN 0370-2693

\bibitem{PhysRevC.79.034909}
B.~I. Abelev, M.~M. Aggarwal, Z.~Ahammed, \emph{et~al.}
\newblock Phys. Rev. C, \textbf{79}:034909 (2009)

\bibitem{STAR:2003fka}
J.~Adams \emph{et~al.}
\newblock Phys. Rev. Lett., \textbf{91}:172302 (2003)

\bibitem{STAR:2006btx}
B.~I. Abelev \emph{et~al.}
\newblock Phys. Rev. Lett., \textbf{98}:192301 (2007).
\newblock [Erratum: Phys.Rev.Lett. 106, 159902 (2011)]

\bibitem{PHENIX:2006wwy}
S.~S. Adler \emph{et~al.}
\newblock Phys. Rev. C, \textbf{76}:034904 (2007)

\bibitem{Adare_2008}
A.~Adare, S.~Afanasiev, C.~Aidala, \emph{et~al.}
\newblock Physical Review Letters, \textbf{101}~(16):162301 (2008)

\bibitem{aliceChargedParticlePP}
B.~Abelev, J.~Adam, D.~Adamov{\'a}, \emph{et~al.}
\newblock The European Physical Journal C, \textbf{73}:1--12 (2013)

\bibitem{ALICEpionetaPP}
S.~Acharya, D.~Adamov{\'a}, M.~M. Aggarwal, \emph{et~al.}
\newblock The European Physical Journal C, \textbf{77}:1--25 (2017)

\bibitem{PhysRevC.101.044907}
S.~Acharya, D.~Adamov\'a, S.~P. Adhya, \emph{et~al.}
\newblock Phys. Rev. C, \textbf{101}:044907 (2020)

\bibitem{CMS:2017uuv}
A.~M. Sirunyan \emph{et~al.}
\newblock Eur. Phys. J. C, \textbf{78}~(6):509 (2018)

\bibitem{2018474}
A.~Sirunyan, A.~Tumasyan, W.~Adam, \emph{et~al.}
\newblock Physics Letters B, \textbf{782}:474--496 (2018).
\newblock ISSN 0370-2693

\bibitem{2022137077}
G.~Aad, B.~Abbott, D.~Abbott, \emph{et~al.}
\newblock Physics Letters B, \textbf{829}:137077 (2022).
\newblock ISSN 0370-2693

\bibitem{TsallisFit}
C.-Y. Wong and G.~Wilk.
\newblock Acta Physica Polonica B, \textbf{43}~(11):2047--2054 (2012)

\bibitem{ALICEChargedParticlePBPB}
B.~Abelev, J.~Adam, D.~Adamová, \emph{et~al.}
\newblock Physics Letters B, \textbf{720}~(1):52--62 (2013).
\newblock ISSN 0370-2693

\bibitem{ALICEpionetaPbPb}
S.~Acharya, F.~T.-. Acosta, D.~Adamov\'a, \emph{et~al.}
\newblock Phys. Rev. C, \textbf{98}:044901 (2018)

\bibitem{das2024charmhadronshothadronic}
S.~K. Das, J.~M. Torres-Rincon, and R.~Rapp.
\newblock Charm and Bottom Hadrons in Hot Hadronic Matter (2024)

\bibitem{Spousta_2016}
M.~Spousta and B.~Cole.
\newblock The European Physical Journal C, \textbf{76}~(2) (2016)

\bibitem{baier2001quenching}
R.~Baier, Y.~L. Dokshitzer, A.~H. Mueller, \emph{et~al.}
\newblock Journal of High Energy Physics, \textbf{2001}~(09):033 (2001)

\bibitem{Buzzatti_2013}
A.~Buzzatti and M.~Gyulassy.
\newblock Nuclear Physics A, \textbf{904–905}:779c–782c (2013).
\newblock ISSN 0375-9474

\bibitem{DOKSHITZER2001199}
Y.~Dokshitzer and D.~Kharzeev.
\newblock Physics Letters B, \textbf{519}~(3):199--206 (2001).
\newblock ISSN 0370-2693

\bibitem{ALICE-deadcone-nature}
S.~Acharya \emph{et~al.}
\newblock Nature, \textbf{605}~(7910):440--446 (2022).
\newblock [Erratum: Nature 607, E22 (2022)]

\bibitem{CMSNonPromptD0}
A.~M. Sirunyan, A.~Tumasyan, W.~Adam, \emph{et~al.}
\newblock Phys. Rev. Lett., \textbf{123}:022001 (2019)

\bibitem{ALICENonPromptD0}
S.~Acharya, D.~Adamová, A.~Adler, \emph{et~al.}
\newblock Journal of High Energy Physics, \textbf{2022} (2022)

\bibitem{Chen_2022}
S.-Y. Chen, J.~Yan, W.~Dai, \emph{et~al.}
\newblock Chinese Physics C, \textbf{46}~(10):104102 (2022)

\bibitem{Song_2017}
L.-H. Song, L.-W. Yan, and C.-G. Duan.
\newblock Chinese Physics C, \textbf{41}~(2):023102 (2017)

\bibitem{PhysRevC.98.024904}
M.~Petrovici, A.~Lindner, A.~Pop, \emph{et~al.}
\newblock Phys. Rev. C, \textbf{98}:024904 (2018)

\bibitem{harris2023qgp}
J.~W. Harris and B.~Müller.
\newblock "QGP Signatures" Revisited (2023)

\bibitem{Zhang_2023}
M.~Zhang, Y.~He, S.~Cao, \emph{et~al.}
\newblock Chinese Physics C, \textbf{47}~(2):024106 (2023)

\bibitem{PhysRevC.111.014907}
J.~Zhao, J.~Aichelin, P.~B. Gossiaux, \emph{et~al.}
\newblock Phys. Rev. C, \textbf{111}:014907 (2025)

\bibitem{ALICE_flow}
R.~Snellings.
\newblock Journal of Physics G: Nuclear and Particle Physics,
  \textbf{41}~(12):124007 (2014)

\bibitem{Chen:2024aom}
J.~Chen \emph{et~al.}
\newblock Nucl. Sci. Tech., \textbf{35}~(12):214 (2024)

\bibitem{PathLengthv2Calc}
B.~Betz and M.~Gyulassy.
\newblock Journal of High Energy Physics, \textbf{2014}~(8):1--25 (2014)

\end{thebibliography}

\end{document}